# Superconducting noise bolometer with microwave bias and readout for array applications

A. A. Kuzmin[1], A. D. Semenov[2], S. V. Shitov[3,4], M. Merker[1], S. H. Wuensch[1], A. V. Ustinov[3,5] and M. Siegel[1]

[1]*Institute of Micro- and Nanoelectronic Systems KIT, 76187 Karlsruhe, Germany*
[2]*DLR Institute of Optical Sensor Systems, 12489 Berlin, Germany*
[3]*National University of Science and Technology MISIS, 119049, Moscow, Russia*
[4]*Kotel'nikov Institute of Radio Engineering and Electronics, 125009, Moscow, Russia*
[5]*Physical Institute, Karlsruhe Institute of Technology KIT, 76131 Karlsruhe, Germany*

We present a superconducting noise bolometer for terahertz radiation, which is suitable for large-format arrays. It is based on an antenna-coupled superconducting micro-bridge embedded in a high-quality factor superconducting resonator for a microwave bias and readout with frequency-division multiplexing in the GHz range. The micro-bridge is kept below its critical temperature and biased with a microwave current of slightly lower amplitude than the critical current of the micro-bridge. The response of the detector is the rate of superconducting fluctuations, which depends exponentially on the concentration of quasiparticles in the micro-bridge. Excess quasiparticles are generated by an incident THz signal. Since the quasiparticle lifetime increases exponentially at lower operation temperature, the noise equivalent power rapidly decreases. This approach allows for large arrays of noise bolometers operating above 1 K with sensitivity, limited by 300-K background noise. Moreover, the response of the bolometer always dominates the noise of the readout due to relatively large amplitude of the bias current. We performed a feasibility study on a proof-of-concept device with a $1.0 \times 0.5$ μm$^2$ micro-bridge from a 9-nm thin Nb film on a sapphire substrate. Having a critical temperature of 5.8 K, it operates at 4.2 K and is biased at the frequency 5.6 GHz. For the quasioptical input at 0.65 THz, we measured the noise equivalent power $\approx 3 \times 10^{-12}$ W/Hz$^{1/2}$, which is close to expectations for this particular device in the noise-response regime.

The demand for sensitive large-format arrays of THz detectors today exists in different terrestrial applications, e.g. THz imaging for non-destructive testing or imaging spectroscopy for material research[1-4]. The array performance, limited only by the noise of the room-temperature background radiation (noise equivalent power, NEP$_{photon}\approx 10^{-15}$ W/Hz$^{1/2}$), is desired at a moderate operation temperature ($T_b>2$ K), which is provided by low-cost and compact cryogenic coolers. Previously, THz imaging arrays with ~$10^2$ pixels were demonstrated[5,6]. They met the sensitivity requirements, but suffered from rapidly growing complexity of the readout with an increase of the array size. A remarkable example of an array-scalable sensor is the microwave kinetic inductance detector (MKID)[7], which is based on high-Q superconducting resonators. Along with frequency-division multiplexing (FDM) in the GHz range, this approach allows for building kilo-pixel arrays using powerful software-defined radio (SDR)[8]. However, THz-range pair-braking MKIDs are using superconductors with small energy gap and operating at $T_b$~100 mK, which requires expensive and bulky cryogenics. Another type of THz-range array-scalable detector, which is able to work at $T_b>2$ K, is the kinetic inductance bolometer (KIBs)[9,10]. Each KIB is a lumped-element resonator on a suspended absorptive membrane. The NEP$\approx 10^{-14}$ W/Hz$^{1/2}$ was reported for the large array of KIBs[11]. However, due to relatively high AC losses, the resonators provided reasonably high quality factors (Q-factor) only at low frequencies in the range of 100 MHz. Along with the more complicated technology of large suspended membranes, low-frequency readout makes it difficult to increase the number of pixels per readout channel.

Here, we propose to combine the advantages of our previously demonstrated superconducting noise bolometer[12] at moderate temperatures and FDM in the GHz range by embedding such a bolometer into a quarter-wavelength (λ/4) superconducting high-Q resonator (Fig. 1 a). The resonator is excited via its shorted end which is weakly coupled to a feed line. General features of the microwave design and the fabrication technology were tested with radio-frequency transition edge sensors (RFTES) [13,14].

Operation conditions of the noise bolometer are: $T_b<T_c$ and $T_b<0.5T_{c,res}$, where $T_c$ and $T_{c,res}$ are the critical temperatures of the micro-bridge and the resonator respectively. The micro-bridge is biased with a microwave signal, applied to port 1 of the feed line (Fig. 1 a). The resonance dips in the transmission spectrum of the feed line $|S_{21}|^2(f)$ (Fig. 1 b) are recorded using a vector network analyzer (VNA) with low-power signal, swept near the resonance frequency $f_{res}$. When the power is below a certain critical value, the resonance dip has a Lorentzian shape. At higher microwave power, when critical current of the micro-bridge is reached, some part of the bridge switches to normal state. If the electron-energy relaxation time in the bridge is longer than $\tau=1/f_{res}$, the resonator enters a regime of steady losses, which causes a reduction of its Q-factor. Losses distort the shape of the resonance dip resulting in the "cratered" Lorentzian curves as it shown in Fig. 1 b. Similar phenomena were described in several works on superconducting resonators with high microwave drive or with embedded "weak points"[15-19]. In a DC-biased noise bolometer, according to the initial concept[20], superconducting fluctuations will lead to random dissipation events in sub-critically biased bridges.

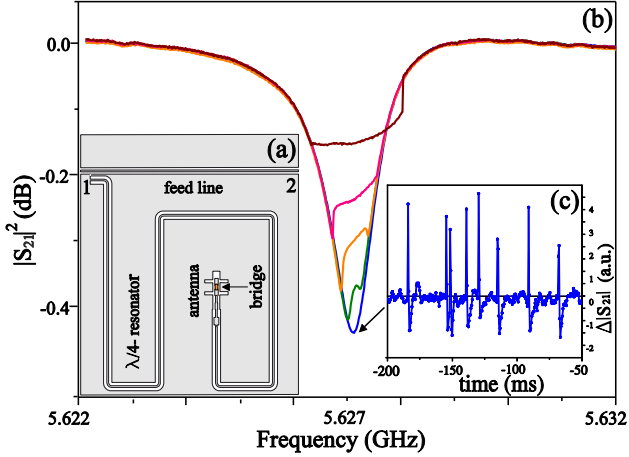

FIG. 1. (a) Schematic layout of the array-scalable noise bolometer. The ports 1 and 2 are on the feed line. (b) Resonance dips in the transmission spectrum for different input powers. (c) The in-phase signal from the IQ-mixer, which represents superconducting fluctuation noise. The noise trace was acquired with a bandwidth 30 kHz – 2 MHz for subcritical microwave bias.

This results in a random sequence of voltage pulses, seen as an excess electric noise. The intensity of this noise serves as the quantity corresponding to a certain level of incident radiation. The magnitude of voltage pulses can be much larger than any noise of the readout, since the non-dissipative bias current in superconducting state of the bridge can be relatively high. By counting voltage pulses with a fast discriminator, one can eliminate noise of the readout completely. In superconducting nanowire single-photon detectors[21], such fluctuation events are known as dark counts and they are characterized by the mean repetition rate $r_0$ – the dark count rate (DCR). Recently, several mechanisms of superconducting fluctuations have been experimentally analysed in mesoscopic superconducting structures[22,23] ($\xi<<w<<\Lambda$, $\xi$ – coherence length, $w$ – width, $\Lambda=2\lambda^2/d$, – Pearl length, $\lambda$ – magnetic penetration depth, $d$ - thickness). The main mechanisms are vortex hopping (VH) over the potential barrier and thermal unbinding of vortex-antivortex pairs (VAP). In both cases DCR at a given temperature $T$ is proportional to the thermodynamic probability:

$$r_0 \propto \exp[-U_b(\Delta, I)/k_B T], \quad (1)$$

where $U_b(\Delta, I)$ is the height of the potential barrier either for VH or for the unbinding of VAP, $\Delta$ – superconducting energy gap, $I$ – bias current. The exact expressions for $U_b$ can be found in Refs. 22 and 24. With the parameters of our micro-bridge and the actual range of bias currents they can be approximated as follows:

$$U_b^{VH} \propto 0.5A\ln(I/I_0)$$
$$U_b^{VAP} \propto A\varepsilon^{-1}\left[(I/2.6I_c) + \ln(2.6I_c/I) - 1\right], \quad (2)$$

where $I_c$ is the critical current, $A = \pi\Phi_0^2\Delta/hR_s$, $R_S$ is the square resistance of the micro-bridge, $\Phi_0$ is the magnetic flux quantum and $\varepsilon$ – the mean polarizability of VAPs, $I_0 = \pi^2\Phi_0\Delta/2hR_s$. The change in the energy gap $\delta\Delta$ due to generation of excess quasiparticles can be estimated from[25]:

$$\delta\Delta/\delta P_{abs} \approx \eta_{pb}\tau_{qp}/(N_0 V \Delta), \quad (3)$$

where $\eta_{pb}$ –the pair-braking efficiency, $\delta P_{abs}$ – the absorbed power, $\tau_{qp}$– quasiparticle lifetime, $N_0$ – density of electron states and $V$ – the volume of the bridge. The internal responsivity $S_{int}$ of the noise bolometer is defined as $S_{int}=\delta r_0/\delta P_{abs}$ and can be estimated from (1)-(3). On the other hand, for independent fluctuation events, the standard deviation in $r_0$ is $\sigma_r\approx(2\pi B r_0)^{1/2}$, where $B$ is the video bandwidth (VBW). Thus internal noise equivalent power $NEP_{int}\equiv\sigma_r/(B^{1/2}S_{int})$ scales as $V/\tau_{qp}$. Since quasiparticle lifetime scales with temperature[26] as $\tau_{qp}\propto\exp(\Delta/kT)/T^{1/2}$, $NEP_{int}$ decreases rapidly with decreasing temperature.

For a noise bolometer integrated with a resonator, dissipation events produce pulsed variations of the transmittance $|S_{21}|$. Here, we will use the terms "dark count" and "DCR" to refer to $|S_{21}|$-pulses and their mean rate $r_0$ respectively. When the superconducting state is fully recovered after a dark count event, the fall time of the $|S_{21}|$-pulse will be $\tau_{fall}=Q_0/\pi f_{res}$, where $Q_0$ is the undisturbed value of the Q-factor. The maximum $r_0$ will be limited by this time. The rise-time $\tau_{rise}=Q_1/\pi f_{res}$ is shorter, since $Q_1<Q_0$ due to losses introduced by dark count event. In order to use a change of DCR as a response, one would need a readout system with a VBW larger than that one of a VNA. In this case the well-developed readout electronics for MKIDs[8], can be used to simultaneously bias and readout multiple noise bolometers through a common feed line. In addition to the advantages of the DC noise bolometer, this approach provides scalability up to arrays with ~$10^3$ pixels biased and read out through a single feed line. Due to the absence of any DC connections, the resonator-coupled noise bolometer is intrinsically well protected from external disturbances and interferences.

We performed a feasibility study at a temperature of 4.2 K using a relatively large single-pixel device. The micro-bridge with a length of 1.15 μm a width of 0.6 μm and a critical temperature of $T_c$=5.8 K was made from a 9-nm thin Nb film, deposited onto sapphire. The λ/4 resonator as well as the feed line was made from a 200-nm Nb film ($T_{c,res}\approx9$ K). The DC properties and superconducting parameters of the micro-bridge were evaluated using an identical witness bridge on the same chip. It has a normal-state resistance $R_n$=69 Ω, a critical current $I_c\approx250$ μA at 4.2 K and a Ginzburg-Landau coherence length of $\xi(0)\approx d\approx10$ nm. Thermal coupling of the micro-bridge to the substrate was estimated from the temperature dependence of the hysteretic current. A Joule power of ≈230 nW was required to keep the witness bridge at $T_c$ for the bath temperature $T_b$=4.2 K. The microwave design of the noise bolometer is similar to the design of RFTES, which was studied earlier[27]. The cratered resonance dips $|S_{21}|^2(f, P)$ were recorded with a VNA (Fig 1. (b)) using a calibrated $^4$He dipstick. At low microwave power, the resonance dip appears at a frequency of $f_{res}\approx5.627$ GHz and has no craters. The loaded quality factor is $Q_L\approx4\times10^3$. Hence, the expected fall time of DCR pulses is $\tau_{fall}$~1 μs. For a 9-nm thick Nb film the electron-energy relaxation time[28] ≈0.5 ns is longer than $1/f_{res}$. Therefore, the Nb micro-bridge introduces steady-state losses in the resonator if the microwave current exceeds the critical current $I_c$. Following this interpretation, we define the critical power of the microwave bias $P_c$ as the power, which corresponds to the onset of the crater in $|S_{21}|^2(f, P)$.

For the measurement of the DCR and THz-radiation response we used the setup schematically shown in Figure 2. The chip with the device is placed onto hyper-hemispherical sapphire lens at a focal point. The output microwave signal is amplified by a cryogenic low-noise amplifier. All components are mounted in a $^4$He bath cryostat with a THz-transparent window. A calibrated 0.65-THz continuous-wave (CW) quasi-optical source was used to illuminate the device. The THz signal was filtered twice, by a mesh filter at 300 K and by a Zitex G106[29] foil at 4.2 K. Since the energy of 0.65-THz photons is larger than $2\Delta$ in the Nb micro-bridge, incident radiation generates excess quasiparticles and suppresses the superconducting energy gap. For the fast $S_{21}$-measurement a homodyne readout scheme was used. It includes a carrier generator, IQ mixer with the 3.5-GHz bandwidth at intermediate frequency and a 100-MHz oscilloscope as the back-end. The Q-component was adjusted to zero with a phase shifter. Any transient in $|S_{21}|$ produce a voltage pulse at the I-port. The shape of the pulse, which represents the dark count, follows the envelope of the transmitted microwave signal. For quantitative analysis of the DCR, a SR620 (*Stanford Research Systems*) pulse counter was used.

A few voltage transients, which represent a variation in $|S_{21}|$ as function of time, are shown in Figure 1 (c). The microwave power was set below the critical value. Random dark counts, caused by superconducting fluctuations in the bridge, are clearly resolved. Measurements with the maximum VBW 30 kHz – 100 MHz showed that all pulses are identical and have a rise-time of $\tau_{rise} \approx 50$ ns and a fall-time of $\tau_{fall} \approx 0.5$ μs. The dependences of both, $r_0$ and $|S_{21}|$, on the microwave power $P_m$ in the form of the relative amplitude of the bias current $I/I_c=(P_m/P_c)^{1/2}$, are shown in Figure 3. Dark counts appear at $I<I_c$ and that $r_0$ increases exponentially with $I/I_c$ as one would expect for the VH or VAP-unbinding events. The DCR saturates at $r_0=r_{max} \approx 5 \times 10^5$ s$^{-1}$, when the amplitude of the current corresponds to the onset of the crater. We invoke the most advanced model[24] of the vortex hopping to fit the hopping rate to our experimental $r_0(I/I_c)$ dependence. The best fit (dashed line in Fig. 3) delivers the penetration depth $\lambda_{fit} = 588$ nm.

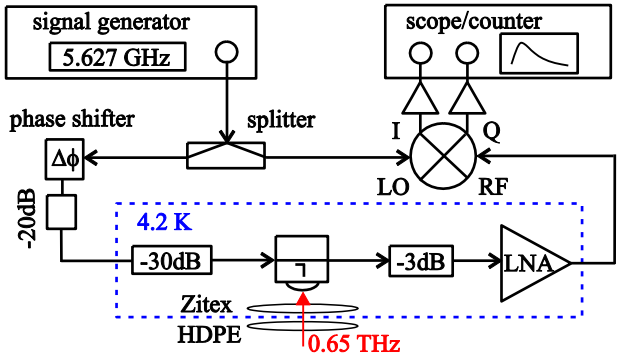

FIG. 2. Scheme of the experimental setup for fast $S_{21}$ measurement.

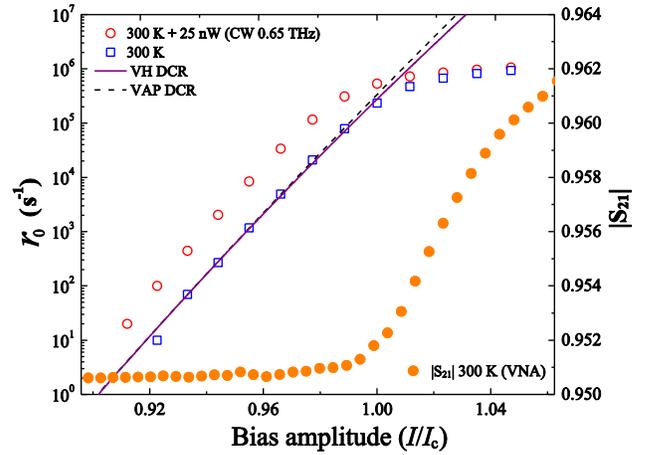

FIG. 3. Transmittance of the feed line (closed symbols) and the dark count rate from the micro-bridge exposed to 300-K background (open squares) and to both 300-K background together with CW 0.65-THz radiation (open circles) at different amplitudes of the microwave current. Lines show the best fit of DCR with the VH model (dashed line) and the VAP model (solid line).

This value is almost twice the value $\lambda_{dc}=283$ nm, which is estimated from DC measurements. The best fit with the VAP model (solid line in Fig. 3) almost coincides with the best fit with the VH model at small currents. In the VAP model the polarizability ε of the VAPs appears as additional fitting parameter. For ε=1 (all pairs are oriented perpendicular to the current), the best fit value is $\lambda_{fit}=471$ nm. For ε=2.8 (averaged polarization of all VAPs 35%) the best fit value of $\lambda_{fit}$ coincides with $\lambda_{dc}$. Since the polarizability of VAPs cannot be assessed independently, we are not able to separate VH and VAP contributions. However, judging from the best fit values and the best-fit lines the VAP scenario seems to be more relevant.

Figure 3 shows that DCR increases when CW 0.65-THz coherent radiation is superimposed on the 300-K background. To obtain device sensitivity to THz waves, we estimated the THz-coupling efficiency through our optics as $\eta=P_J/P_S \approx 5\%$, where $P_J$ is the Joule power of the DC hysteretic current and $P_S$ is the THz power, which suppresses completely the resonance dip in $|S21|$ at the microwave power corresponding to $I/I_c<10^{-3}$. The low value of η is mostly due to quasi-optical beam mismatch between the sapphire lens and the horn antenna of the THz source. The operation point $0.98I/I_c$, ($r_{bias} \approx 2 \times 10^4$ s$^{-1}$) provided the highest optical responsivity $S_{THz}=\delta r_0/\delta P_{THz}$ for a small THz excitation $P_{THz} \ll P_s$ ($P_{THz} \approx 20$ nW). For this operation point the internal noise equivalent power was estimated as

$$\text{NEP}_{int} \approx \eta \sigma_r / (\sqrt{B} \cdot S_{THz}) \qquad (4)$$

resulting in $\approx 3 \times 10^{-12}$ W/Hz$^{1/2}$ at 4.2 K for the used $B=1$ Hz. The dependence of $r_0$ on the absorbed THz power $\eta P_{THz}$ at the optimal bias is shown in Figure 4 along with the power dependence of the responsivity. For the optimal bias, saturation in responsivity begins at $P_{sat} \approx 10$ nW, which defines a dynamic range DR=10log$_{10}[P_{sat}/\text{NEP}_{int}B^{1/2}] \approx 35$ dB can be estimated for $B=1$ Hz (or DR $\approx 10\log_{10}[r_{max}/(Br_{bias})^{1/2}]$).

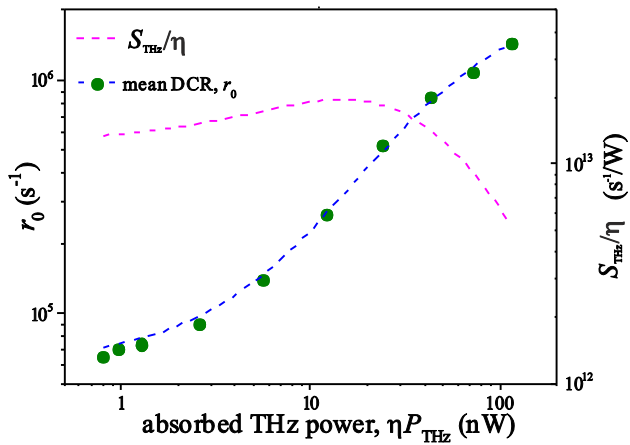

FIG. 4. Dependences of the mean DCR (dashed line is the fit) and the responsivity for the optimal bias point.

It is worth to note that we did not take any special measures to stabilize the temperature in the cryostat or to avoid vibrations of the flexible cables, which lead to excess $1/f$-noise.

In conclusion, direct time-resolved $S_{21}$-measurements of the resonator-coupled micro-bridge revealed expected excess noise in the form of $|S_{21}|$-transients, which appear due to fluctuations in the superconducting state. The measured current dependences of the fluctuation rate support theoretical vortex-based scenario of the noise origin. At 4.2 K optically measured NEP of the proof-of-concept device with the relatively large micro-bridge is in agreement with the estimations in the framework of the noise-bolometer model[18]. Since the NEP should scale with the bath temperature as $T^{1/2}\exp(-\Delta/k_B T)$, the value $10^{-14}$ W/Hz$^{1/2}$ is expected at $T_b \approx 1.5$ K even for the tested large device. This sensitivity is in accordance with the NEP of the RFTES which was measured previously with a matched cryogenic blackbody[26]. This RFTES had similar size and was operated at 1.5 K in the cratered regime of the resonator. In order to reach NEP$\approx 10^{-15}$ W/Hz$^{1/2}$ at $T_b \approx 4$ K, the micro-bridge should be sized down to the sub-micron range that is readily achievable with e-beam lithography. Additionally, the substrate under the micro-bridge could be etched away. In this case the effective quasiparticle lifetime and thus the sensitivity will be increased due to trapping of 2Δ-phonons inside the micro-bridge[30]. This improvement is rather easy to implement, since for a small bridge there is no need for large-area suspended membranes[31, 32].

This work was supported in parts by grant 12-02-01352-a from Russian Foundation for Basic Research; Increase Competitiveness Program of NUST «MISiS» (№ K2-2016-051) and contract 11.G34.31.0062 from the Ministry of Education and Science of the Russian Federation; grants 05K13VK4, 13N12025 from German Federal Ministry of Education and Research (BMBF) and project № 284456 of the European Commission.